# Nonequilibrium electron spin polarization in a double quantum dot. Lande mechanism.

Yuri A. Serebrennikov

Qubit Technology Center

2152 Merokee Dr., Merrick, NY 11566

ys.qubit@gmail.com

In moderately strong magnetic fields, the difference in Lande *g*-factors in each of the dots of a two-electron double quantum dot (DQD) may induce oscillations between singlet and triplet states of an entangled pair and lead to a nonequilibrium electron spin polarization in the system. The calculated bipolar polarization reflects the strong spin correlation in the spatially separated pair of electrons that occupy almost degenerate orbital states in a coupled DQD. We will show that this polarization may partially survive the rapid inhomogeneous decoherence caused by random nuclear magnetic fields. 73.21.La, 76.30.-v

In recent years, beautiful and convincing experiments demonstrate that semiconductor quantum dots (QD's) provide the ultimate level of control and measurement over the single electron charge and spin[1]. The spin selection rules strongly influence the low temperature transport and may completely block the current trough the dot. Situation becomes especially interesting in a coupled double quantum dot (DQD) system with two confined carriers where electron spins are entangled in singlet or triplet states that can be controlled by electric gates and external magnetic fields[2]. These results open new opportunities for the design of solid-state quantum computers where the spin rather than the charge of an electron is used for information processing and storage. Besides its importance in technology, ability to control single carrier confinement energy, a number of interacting particles, as well as the carrier-carrier exchange interaction is of considerable interest for our understanding of the fundamental properties of spin-correlated fermionic systems.

Conservation of spin during a tunneling event blocks the transitions from the (1,1) triplet states to the (0,2) singlet state of an entangled two-electron system in a DQD, where (*n,m*) denotes the number of electrons in each dot. Since the (0,2) triplet configuration is significantly above the (0,2) ground singlet state (~ 0.4 meV >> $kT$ at T ~1 K), transitions between almost degenerate triplet and singlet states in the (1,1) configuration determine the charge transport through the DQD device[2]. In III-V semiconductors, the main driving force for these transitions in zero and small magnetic fields ($B$ < 10 mT) is a contact hyperfine coupling (HFC) with lattice nuclei[2,3,4,5,6], which may influence the spin dynamics of electrons confined within a DQD, by mixing the singlet $|S>$ and triplet $|T_{0,\pm}>$ states of a (1,1) spin correlated pair. The sum of an external magnetic field $B$ and nuclear hyperfine fields, resulting from the various nuclear spins, determines the precession frequency of the individual electron spin in the QD. In general, the direction and strength of the resulting magnetic field is different in each dot. This difference is the actual source of the singlet-triplet mixing and quantum beats in the (1,1) configuration of a DQD [2,3]. The large number of nuclear spins in, e.g., GaAs quantum dot (N ~ $10^6$) leads, however, to a very complex electron spin dynamics and fast inhomogeneous spin dephasing ~ 10 ns (see [3][4][5][6] and references to the earlier literature therein).

Precession frequency difference at two spatially separated spin sites and the related singlet-triplet mixing and transitions may also arise from the difference in Lande $g$-factors. This mechanism was proposed more then 50 years ago to explain the magnetic field effect on the ortho-para conversion in positronium[7]. The large difference in g-factors of electron and positron, $g_e - g_p = 4$, leads to the strong coupling of the $F_Z = 0$ component of the orto-state (the total spin of the orto-positronium, $F = 1$) and para-state ($F = 0$) in a moderate magnetic fields ($B < 1$ T), thereby opening up an efficient two-photon annihilation channel for orto-positronium. The physical principle of this mechanism was rediscovered in late sixties in the context of spin-dependent radical reactions ("spin chemistry")[8]. Since then, the spin physics was actively employed to design or interpret magnetic field effects on chemical reactions. In particular, it has been well established that in the presence of the Heisenberg exchange interaction the difference $\Delta g$ in electron $g$-factors in a spin correlated radical pair may give rise to a nonequilibrium spin polarization in a system[8]. The basic physics of the phenomenon is very general. Therefore, one may expect that the same mechanism of magnetic field effects could be operative in a near-degenerate singlet and triplet states (1,1) of a two-electron DQD, which can be considered as an artificial bi-radical.

Usually for organic radicals $\Delta g$ is very small $\sim 10^{-2} - 10^{-3}$. It is well known, however, that in semiconductor nanostructures $g$-factor strongly depend on the spatial confinement, size, shape, and material of the QD; external electric field, applied stress, quantum well width $W$, and the inter-well coupling[9][10][11]. For example, the measured difference between g factors in GaAs quantum wells[9] $\Delta g = g(W = 10 \text{ nm}) - g(W = 7\text{nm})$ = 0.14. In this case, increasing linearly with an external magnetic field, the difference in Larmor frequencies $2\pi \Delta \nu_L = 0.14 \beta B / \hbar$ will become larger than the precession frequency of an electron spin in an effective hyperfine field in GaAs QD at fields $B \sim 50$ mT. The main question to be addressed in III-V semiconductors is the time behavior and the magnitude of the effect(s) in the presence of random hyperfine fields that will certainly result in a decay of singlet-triplet oscillations and will act to destroy a nonequilibrium electron spin polarization in the system. Yet, it is not clear how far will this process go in a *single* two-electron DQD. Here we will show that in moderately strong magnetic fields Lande mechanism (referred also as "$\Delta g$ mechanism"[8]) of magnetic field effects yields nonequilibrium electron spin polarization in each dot that will partially survive the rapid inhomogeneous decoherence.

The effective spin Hamiltonian of a coupled two-electron (1,1) system interacting with nuclear spins in a DQD via the contact HFC has been recently derived by Coish and Loss[6]. The straightforward generalization of this Hamiltonian yields

$$H = \vec{h} \cdot \vec{S} + \delta\vec{h} \cdot \delta\vec{S} + (J/2)\vec{S} \cdot \vec{S} - J \tag{1}$$

Here $\vec{S} = \vec{S}_1 + \vec{S}_2$, $\delta\vec{S} = \vec{S}_1 - \vec{S}_2$, $\vec{h} = (\vec{h}_1 + \vec{h}_2)/2$, $\delta\vec{h} = (\vec{h}_1 - \vec{h}_2)/2$, and

$$\vec{h}_i = g_i \beta \vec{B} + (\sum_{j \in i} A_j \vec{I}_j)_i, \tag{2}$$

where for simplicity $g_i$ is assumed to be isotropic. Index $i = 1, 2$ enumerates the quantum dot, $\beta$ is the Bohr magneton, $\vec{S}_i$ and $\vec{I}_j$ are the electron and nuclear spin operators in each dot, $A_j$ is the HFC constant at the lattice site $j$, notation $j \in i$ is shorthand for all nuclei that belong to the dot $i$, and $J$ is the scalar exchange coupling constant. The electron spin of each member of a spin-correlated pair is precessing in a different local magnetic field $\vec{h}_i / g_i \beta$. As time goes on, the difference in precessional frequencies $\Delta \nu_L \sim |\vec{\delta h}|$ leads to a change in the total spin of an entangled two-electron system $S^2 = (\vec{S}_1 + \vec{S}_2)^2$ and, hence, to dynamic singlet-triplet conversion[3,4,5,6,7,8]. In zero external magnetic field any transitions between the spin substates of the (1,1) electron pair induced by the difference in hyperfine fields are possible. Note, however, that as the external magnetic field increases, the precession axes of the two spins become increasingly collinear (the projection of the total electron spin operator on the direction of the field turns into a good quantum number), suppressing transitions between the (1,1) $|S>$, $|T_0>$ and $|T_\pm>$ eigenstates when the electron Zeeman energy dominates the HFC. In sufficiently large magnetic field $B > 10$ mT, which we will presume to be parallel to a DQD it is convenient to choose the direction of $\vec{B}$ as the quantization axis $Z$ of the spin operators. In this "high field limit", taking into account the assumed isotropy of the Heisenberg exchange interaction and dropping the constant term in Eq.(1), we obtain[6,8]

$$\tilde{H} = (J/2) S^2 + h_z \cdot S_z + \delta h_z \cdot \delta S_z, \qquad (3)$$

where $\delta h_z = \delta h_z^e + \delta h_z^n$,

$$2\delta h_z^e = (g_1 - g_2) \beta B, \qquad 2\delta h_z^n = \sum_{j \in 1} A_j m_j - \sum_{j \in 2} A_j m_j,$$

and $m_j$ is the corresponding nuclear spin quantum number.

The isotropic exchange term determines the singlet-triplet splitting at $B = 0$. The second term also conserves $S^2$ and $S_z$, and is responsible for the synchronous precession of the individual spins about the direction of the external magnetic field. The third term of the high field Hamiltonian does not commute with $S^2$, thereby giving rise to singlet-triplet mixing. However, it is easy to check that in the high field limit $B >> \delta h_z^n / g\beta$, $[\tilde{H}, S_z] = 0$ and, as expected, the Z-component of the total electron spin operator is conserved. Certainly, conservation of $S_z$ does not mean that $S_{1z}$ and $S_{2z}$ are preserved. Due to combined action of the exchange coupling and the difference in electron Larmor frequencies between dots, the Z-component of the i-th spin experiences a nutation that is strongly correlated with its spatially separated partner localized in the adjacent dot (spins of an entangled pair are "waltzing" together).

Projection onto the complex two-dimensional Hilbert space spanned by $S_z = 0$ states yields the Schrödinger-like equation (see Ref.[6] for details)

$$i\dot{\Psi}_{1/2}(t) = H_{eff} \Psi_{1/2}(t), \qquad (4)$$

with the effective spin-Hamiltonian

$$H_{eff} = (J/2)(1+\sigma_z) + \delta h_z \cdot \sigma_x := \vec{\omega} \cdot \vec{S}_{eff} \qquad (5)$$

which governs the evolution of a two-component spinor $\Psi_{1/2}$ that represents the $S-T_0$ doublet, adiabatically isolated at $|\vec{h}| >> kT$ from the rest of the spin-multiplet. Here and in the following $\hbar = 1$, $\vec{\sigma} = \{\sigma_1, \sigma_2, \sigma_3\}$ is the vector of Pauli matrices, and we introduce $\vec{\omega} := \{2\delta h_z, 0, J\}$. Clearly, the effective spin-Hamiltonian $H_{eff}$, Eq.(5), can be viewed as a generic Zeeman Hamiltonian of a spin-1/2 particle that gives rise to precession of the polarization vector $\vec{p}(t) := Tr[\rho(t)\vec{\sigma}]$ in the real 3D space ($\rho(t) := |\Psi_{1/2}(t)\rangle\langle\Psi_{1/2}(t)|$ is the density operator of the $S-T_0$ doublet). Note that the quantization axis of the effective spin operator $\vec{S}_{eff} := \vec{\sigma}/2$ does not coincide with the direction of an external magnetic field: $|\sigma_3 = -1>$ and $|\sigma_3 = 1>$ correspond to $|S>$ and $|T_0>$ respectively (it is supposed that $J > 0$). An exchange interaction drives precession of $\vec{p}$ about the 3-axis, $\omega_3 = J$, while $2\delta h_z$ is equal to the $\omega_1$ component of the effective "magnetic field" $\vec{\omega}$, and determines the nutation of the polarization vector. If $\vec{\omega}(t) = \vec{\omega}$, utilizing the homomorphism between vectors and rotation operators of Euclidean space and spinors and rotation operators of 2D spinor space[12][13][14], we may cast the solution of Eq.(4) into the following form

$$\vec{p}(t) = \vec{p}\cos\phi - (\hat{\vec{n}} \times \vec{p})\sin\phi + \hat{\vec{n}}(\hat{\vec{n}} \cdot \vec{p})(1 - \cos\phi) = \Re_{eff}(\phi, \hat{\vec{n}})\ \vec{p}. \tag{6}$$

Here $\phi := \omega t = [\omega_1^2 + \omega_3^2]^{1/2} t = [4\delta h_z^2 + J^2]^{1/2} t$ denotes the angle of rotation of an effective spin about the axis defined by the unit vector $\hat{\vec{n}} = \{\sin\theta, 0, \cos\theta\}$, $tg\theta = \omega_1/\omega_3$, $\vec{p}$ denotes $\vec{p}(t=0)$, and $\Re_{eff}$ stands for the 3D matrix

$$\Re_{eff}(\phi,\hat{\vec{n}}) = \begin{bmatrix} \sin^2\theta + \cos^2\theta\cos\phi & -\sin\phi\cos\theta & 1/2(1-\cos\phi)\sin 2\theta \\ \sin\phi\cos\theta & \cos\phi & -\sin\phi\sin\theta \\ 1/2(1-\cos\phi)\sin 2\theta & \sin\phi\sin\theta & \cos^2\theta + \sin^2\theta\cos\phi \end{bmatrix} \tag{7}$$

that describes the time evolution of the polarization vector and corresponds to the simple geometrical operation. It represents the covariant rotational transformation of a classical vector, which, regardless of the physical context, is essentially a pure geometric action. In fact, in rigid-body mechanics formula (6) is well known as Rodrigues' formula[15].

Our goal is to describe the time behavior of the probability, $^S w_T(t)$, to find electron spins in the $|T_0>$ state (denoted by the subscript T) if the two-electron system was in the singlet state at time $t = 0$ (the left superscript S), and to calculate the electron spin polarization in each dot, $\mu_{iz}^{(e)}$. These quantities can be readily expressed through the components of $\vec{p}$:

$$^S w_T(t) = 1 - \,^S w_S(t) = [1 + \rho_{T_0 T_0}(t) - \rho_{SS}(t)]/2 = [1 + p_3(t)]/2, \qquad (8)$$

$$-\mu_{1z}^{(e)}(t) := 2Tr[\rho(t) S_{1z}] = Tr[\rho(t)(S_z + \delta S_z)] = p_1(t), \qquad (9)$$

$$\mu_{1z}^{(e)}(t) = -\mu_{2z}^{(e)}(t). \qquad (10)$$

Hence, taking into account Eqs.(6) and (7), for $\vec{p} = \{0, 0, -1\}$ we obtain

$$^S w_T(t) = \sin^2(\phi/2) \sin^2 \theta = \frac{\delta h_z^2}{\delta h_z^2 + J^2/4} \sin^2(t\sqrt{\delta h_z^2 + J^2/4}), \qquad (11)$$

$$\mu_{1z}^{(e)}(t) = \sin^2(\phi/2) \sin(2\theta) = \frac{2J\delta h_z}{\delta h_z^2 + J^2/4} \sin^2(t\sqrt{\delta h_z^2 + J^2/4}). \qquad (12)$$

These formulas can be also derived from the simple geometry of the problem[12]. Indeed, if $\alpha$ is the angle between $\vec{p}(t)$ and $\vec{p}$, then $p_3(t) = -\cos\alpha$, $p_1(t) = \sin\alpha$ (we would like to remind the reader that, in general, the {x,y,z} and {1,2,3} axes do not coincide). Taking into account that $\cos\alpha = \cos^2\theta + \sin^2\theta \cos\phi = 1 - 2\sin^2\theta \sin^2(\phi/2)$ and $\sin\alpha = \sin(2\theta) \sin^2(\phi/2)$ we obtain Eqs.(11) and (12).

Now we are prepared to make a next step and perform the average $<>$ of the $^S w_T(t)$ and $\mu_{1z}^{(e)}(t)$ over the random ensemble of nuclear spin states. We will follow the analysis of Coish and Loss[6] and assume a *quasistatic* continuous Gaussian distribution of $\delta h_z^n$ values with $<\delta h_z^n> = 0$. After minor rearrangement Eqs.(11) and (12) take the form

$$<^S w_T(t)> = <\frac{2\delta h_z^2}{4\delta h_z^2 + J^2}[1 - \cos(t\sqrt{4\delta h_z^2 + J^2}\,)]>, \qquad (13)$$

$$<\mu_{1z}^{(e)}(t)>=<\frac{4J\delta h_z}{4\delta h_z^2+J^2}[1-\cos(t\sqrt{4\delta h_z^2+J^2}\,)]>. \qquad (14)$$

If $\delta h_z^e \ll \delta h_z^n$ we may neglect the difference in electron g-factors, then Eq.(13) recovers the relevant expression obtained in Ref.[6] and $<\mu_{iz}^{(e)}(t)>=0$. In the opposite limit $\delta h_z^e \gg \delta h_z^n$, direct integration of Eqs. (13) and (14) yields

$$<^S w_T(t)> = \frac{2(\delta h_z^e)^2}{4(\delta h_z^e)^2+J^2}[1-\cos(t\sqrt{4(\delta h_z^e)^2+J^2}\,)e^{-(t/t_0)^2/2}], \qquad (15)$$

$$<\mu_{1z}^{(e)}(t)>=-<\mu_{2z}^{(e)}(t)>=\frac{4J\delta h_z^e}{4(\delta h_z^e)^2+J^2}[1-\cos(t\sqrt{4(\delta h_z^e)^2+J^2}\,)e^{-(t/t_0)^2/2}], \qquad (16)$$

where we introduce

$$t_0 = \frac{\sqrt{[4(\delta h_z^e)^2+J^2]/<(\delta h_z^n)^2>}}{4\delta h_z^e}. \qquad (17)$$

Clearly, the physical meaning of this parameter is an effective time of decoherence[6]. Note that with the replacement $\delta h_z^e \to <\delta h_z^n> \neq 0$, Eq.(15) recovers the expression for singlet-triplet interconversion in a DQD in the case of a nonequilibrium inhomogeneous nuclear polarization obtained by Coish and Loss. What is new here is (i) the dependence of $<^S w_T(t)>$ and $t_0$ on the strength of external magnetic field and the difference in g-factors between dots; (ii) the emergence of a nonequilibrium electron spin polarization $<\mu_{iz}^{(e)}(t)>$ in each dot.

Very recently, the *symmetric* ($g_1 = g_2 = g$) DQD device was used to measure the probability $<^S w_S(t)>$ as a function of a gate-voltage tunable exchange interaction in the system[16]. It has been shown that the predictions of the theory[6] are in a good agreement with experimental data. The lifetime of a two-electron singlet state spans over hundreds of ns and $<^S w_S(t)>$ remains close to unity when the exchange interaction is tuned to be larger than the average expectation value of the operator $\delta h_z^n$. In this regime, the rapid precession of $\vec{p}$ about the 3-axis makes hyperfine fields acting in the transverse direction along 1-axis less effective, thereby suppressing singlet-triplet transitions and hyperfine dephasing. On the other hand, according to Lande mechanism, see Eq.(15), the rate of $|S> - |T_0>$ transitions gradually increases with B and one may expect that $<^S w_S(t)>$ and $<^S w_T(t)>$ will saturate at ½ in sufficiently large fields, $4(\delta h_z^e)^2 >> J^2$ when $t >> t_0$. It is well known that an electron g-factor presents strong variations (even changes of sign) with the size and shape of the quantum dot[9][10][11]. Therefore, the large difference in g-factors can be straightforwardly achieved and this condition could be easily satisfied in a weakly coupled *asymmetric* DQD device. In this situation, the polarization vector is precessing with a frequency $\Delta \nu_L$ about the 1-axis, which corresponds to rapid $|S> - |T_0>$ transitions. The hyperfine perturbation $\delta h_z^n$ operates along the same axis and, hence, is responsible for the quasistatic distribution of the precession frequencies. As a result, when $4(\delta h_z^e)^2 >> J^2, <(\delta h_z^n)^2>$ the effective time of decoherence Eq.(17) is approaching the asymptotic value $2t_0 = 1/\sqrt{<(\delta h_z^n)^2>}$. In Figure 1 we plot the behavior of $<^S w_T(t)>$, as given in Eq.(15), for different values of $b = 2\delta h_z^e = (g_1 - g_2)\beta B$ and ratios between $J$ and $\sqrt{<(\delta h_z^n)^2>} = 50$ neV. The latter value corresponds to the hyperfine energies recently obtained in the symmetric GaAs double dot device[2][16]. The figure illustrates that the triplet occupation probability undergoes damped oscillations controlled by the ratio between exchange and hyperfine interactions. Evidently, even a small asymmetry of the device $|\Delta g|=|g_1-g_2|=10^{-2}$ leads to a significant probability to find the singlet-born (1,1) pair in the triplet state at $B$ = 200 mT, which corresponds to $b$ = 2 mT or 0.1 $\mu eV$.

Remarkably, Eq.(16) predicts an emergence of a nonequilibrium electron spin polarization in each of the QD's of an asymmetric DQD device that will survive even at $t \gg t_0$, see Figure 2. The long time asymptotic of this polarization exhibits the clear maximum $<\mu_{1z}^{(e)}(t)>=-<\mu_{2z}^{(e)}(t)>=1$ at $(\Delta g \beta B)^2 = J^2$, which corresponds to $<^S w_T>=1/4$ ($\omega_1 = \omega_3, \theta = \pi/4$) and, therefore, can be rather strong at sub-Kelvin temperatures. We illustrate this behavior in Figure 3. The plot clearly shows that measurements of the electron spin polarization as a function of an external magnetic field yield direct information about the exchange interaction in the DQD. Furthermore, it is easy to see that application of strong fields $(\Delta g \beta B)^2 \gg J^2$ may reduce the magnitude of the effect $<\mu_{iz}^{(e)}(t)> \sim J/\beta B$. The "+/-" polarization pattern predicted by Eq.(16) might be considered as the "fingerprint" of Lande mechanism of magnetic field effect in a spatially separated spin-correlated electron pair[8]. Note that the growth of an exchange interaction in a coupled DQD will increase the effective time of decoherence $t_0$, but may result in diminishing of the electron spin polarization $<\mu_{iz}^{(e)}(t)> \sim \beta B/J$ for $(\Delta g \beta B)^2 \ll J^2$. Likewise, one cannot expect to observe strong polarization in the opposite limit of extremely weak coupling.

In summary, in moderately strong external magnetic fields the interplay between Zeeman and exchange interaction may lead to a nonequilibrium electron spin polarization in each of the dots in the (1,1) configuration of the weakly coupled asymmetric ($\Delta g \neq 0$)

DQD device. This polarization is resilient against inhomogeneous decoherence caused by random nuclear magnetic fields of a lattice (will survive at $t \gg t_0$). The bipolar "+/-" polarization in a coupled DQD reflects the strong spin entanglement in the $|T_0>$ state of a spatially separated pair of electrons that occupy two different but almost degenerate orbital states. Even a small difference in *g*-factors ~ 0.01 can cause a drastic change in the evolution of the singlet and triplet correlators and may lead to a significant probability to find the singlet-born (1,1) pair in the triplet state at moderate fields ~ 200 mT. Consequently, an ability to control the difference in g-factors between the dots by external electric fields or applied stress ("g-factor engineering", see Refs.[10, 11]) provides a novel way to manipulate the spin entanglement in the system. It should be qualitatively clear that if one applies a nonadiabatic pulse in voltage bias emitting electron(s) out of the device, the residual polarization of the spin entangled pair created in the coupled DQD will be transferred into the lead(s).

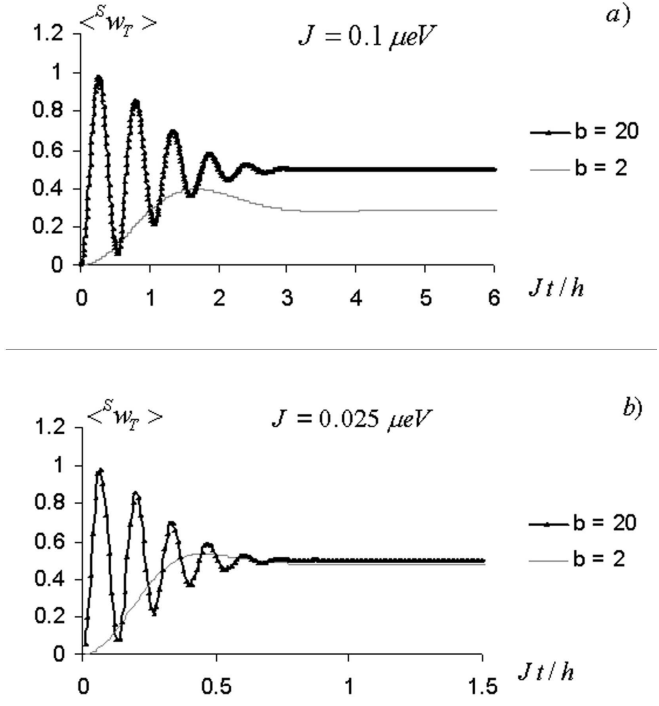

Fig.1 The triplet occupation probability from Eq.(15) for two values of the parameter $b$ mT (see text) that represents the difference in electron Zeeman energies between dots in the asymmetric DQD device. (a) $J = 2\sqrt{<(\delta h_z^n)^2>}$ ; (b) $J = \sqrt{<(\delta h_z^n)^2>}/2$

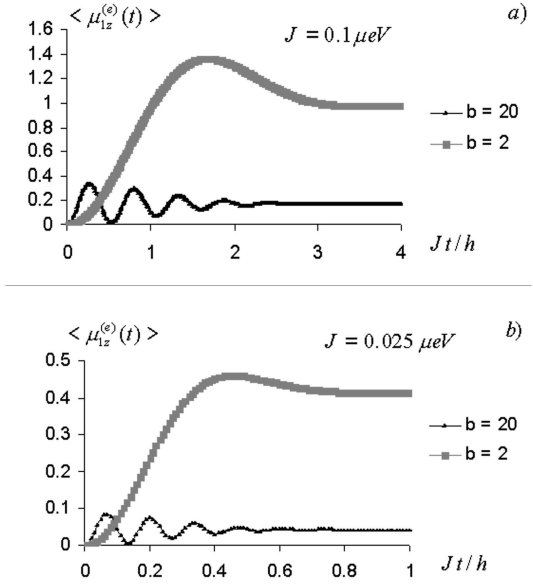

Fig.2 The electron spin polarization from Eq.(16) for two values of the parameter *b* mT (see text) that represents the difference in electron Zeeman energies between dots in the asymmetric DQD device. (a) $J = 2\sqrt{<(\delta h_z^n)^2>}$ ; (b) $J = \sqrt{<(\delta h_z^n)^2>}/2$

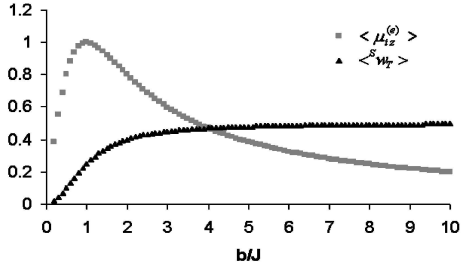

Fig.3 The long time asymptotic value of the triplet occupation probability and electron spin polarization vs. *b/J* from Eqs.(15) and (16) in the asymmetric DQD device.

---

[1] D. Weinman, W. Haussler, and B. Kramer, Phys. Rev. Lett. **74**, 984 (1995); K. Ono, D. G. Austing, Y. Tokura, and S. Tarucha, Science **297**, 1313 (2002); T. Fujisawa, D. G. Austing, Y. Tokura, Y. Hyrayama, and S. Tarucha, Nature **419**, 278 (2002); T. Hayashi,